\documentclass[12pt,preprint]{aastex}

\shorttitle{Size of Hot Gas in Ellipticals}
\shortauthors{Bregman et al.}

\begin{document}


\title{The Size of the Cooling Region of Hot Gas in Two Elliptical Galaxies}

\author{Joel N. Bregman, Birgit Otte, Eric D. Miller\altaffilmark{1}, 
and Jimmy A. Irwin}
\affil{Department of Astronomy, University of Michigan, Ann Arbor, MI 48109}
\email{jbregman@umich.edu, otteb@umich.edu, milleric@mit.edu, jairwin@umich.edu}

\altaffiltext{1}{present address: Kavli Institute for Astrophysics and Space Science, MIT, Cambridge, MA 02139}

\begin{abstract}

Some early-type galaxies show \ion{O}{6} emission, a tracer of gas at
10$^{{\rm 5.5}}$ K, and a predicted product of gas cooling from the
X-ray emitting temperatures.  We studied the spatial extent and velocity
structure of this cooling gas by obtaining spectra of the \ion{O}{6} doublet
in NGC 4636 and NGC 5846 with the {\it Far Ultraviolet Spectroscopic Explorer\/}.  
For NGC 4636, the central LWRS pointing shows that the \ion{O}{6} lines 
are double-peaked and symmetrical about the systemic velocity of the 
galaxy, with a separation of 210 km  s$^{-1}$.  A LWRS observation 30\arcsec \ 
from the center failed to show additional \ion{O}{6} emission.  
For NGC 5846, three spectra were obtained with 4\arcsec\ $\times$ 20\arcsec \ 
apertures (MDRS) at the center and 4\arcsec \ to the east and west of the center.  
The \ion{O}{6} line flux seen in the previous LWRS is contained in
the sum of the smaller apertures, with most of the flux in a single non-central
MDRS aperture, suggesting a size for the emission $\le$ 0.5 kpc; the
emission consists of a blue and red peak.
For both galaxies, the \ion{O}{6} velocity structure is similar to that of 
the optical [NII] emission and is consistent with rotation.
The compactness and velocity structure of the \ion{O}{6} emission rules
out cooling flow models with broadly distributed mass drop-out but is
consistent with cooling flow models where the cooling occurs primarily in
the central region.  The 10$^4$ K gas may be the end state of
the \ion{O}{6} emitting gas.
\end{abstract}

\keywords{galaxies: individual (\object{NGC 4636}, \object{NGC 5846})
--- galaxies: ISM --- cooling flows --- X-rays: galaxies}

\section{Introduction}

Most early-type galaxies have hot X-ray emitting gas, and from
the density and temperature of the gas (10$^{{\rm 7}}$ K), the cooling time is much
less than a Hubble time in the central 10 kpc (review in \citealt{math03}).  The X-ray observations
led investigators to develop a model where the gas cools into neutral or
warm ionized material, possibly forming stars as the endpoint of the
process.  This cooling flow model of early-type galaxies has met with
mixed success.  On one hand, the mass loss from the stars is observed
directly \citep{athey02}, and at least some of this mass loss thermalizes and becomes the
hot X-ray emitting gas.  It was expected that the gas would
radiatively cool yet the expected multi-temperature gas is not seen at X-ray 
wavelengths \citep{xu02}.  However, from ultraviolet line emission studies, and in
particular \ion{O}{6}, there is evidence for gas having cooled below the
temperature of the ambient hot medium \citep{breg01}.  We analyzed a sample of 24
nearby early-type galaxies for their X-ray emission as well as their \ion{O}{6}
emission (using the {\it Far Ultraviolet Spectroscopic Explorer\/}) and found
that more than a third of the galaxies had \ion{O}{6} emission \citep{breg05}.  The \ion{O}{6} ion
has an ionization potential of 113.9 eV, making it difficult to produce
through photoionization.  It is almost certainly ionized by collisional
processes and it is only common at gas temperatures near 10$^{{\rm 5.5}}$ K, about
a factor of 30 below the temperature of the ambient medium.

Models suggest that as the gas radiatively cools, it can undergo
thermal instabilities and the nature of the instabilities determines where
the gas cools within the galaxy.  Initially, it was thought that thermal
instabilities would grow over a wide range of radii (e.g., \citealt{math78}) 
but subsequent calculations showed that the inflow of the gas suppresses 
the growth rate of radial modes, although the instability can occur for
some non-radial modes and modes involving imbedded magnetic fields \citep{balb88,balb89,balb91}.  
If the growth of thermal instabilities is supressed, the gas
will cool primarily in the central region of the galaxies. 
If enough thermal instabilities can grow, or if nonlinear perturbations are present, 
caused by processes such as the mass loss around individual stars,
the cooling of the X-ray emitting gas may be more distributed.

The distribution of the cooling gas can be parameterized within
hydrodynamic models, such as the time-independent models of \citet{sarz89},
where the local mass dropout rate (the conversion of
hot to cool gas that no longer contributes significantly to buoyancy) is
$\dot{\rho} = q{\rho}/t_{cool}$.   In this parameterization, $q = 0$ corresponds
to no mass drop-out, $q = 1$ to a mass loss rate given by the instantaneous
cooling time, and $q = 4$ to a more rapid rate of cooling and mass drop-out.  
The weak growth of thermal instabilities would suggest a value for $q$ near zero, 
while efficient thermal instability growth (linear and non-linear perturbations) would
correspond to q of unity.  The different choices of $q$ lead to different
X-ray surface brightness profiles and luminosities, with low values of $q$
leading to higher total X-ray luminosities and more sharply peaked X-ray
profiles.  Based on the observations at that time, \citet{sarz89} argued for 
models where $q = 1-3$ rather than low values.

The value of $q$ affects both the spatial distribution of the cooling
gas as well as its velocity, with low values of $q$ corresponding to larger
flow velocities to satisfy the mass conservation equation.  At a radius of
0.5 kpc, the inflow velocities are 15 km s$^{-1}$ for $q = 4$ and 40 km s$^{-1}$ for $q
= 0$ \citep{sarz89}.  Whereas such velocities cannot be observed with current X-ray
instrumentation, it is accessible through observations of the \ion{O}{6}
$\lambda$$\lambda$1032,1038 lines, where the resolution of {\it FUSE\/} is 15 km s$^{-1}$. 
Therefore, we have used {\it FUSE\/} to examine both the line width and
spatial distribution of the \ion{O}{6} line for two early-type galaxies that we
detected in our larger survey, NGC 4636 and NGC 5846.  Previously we
used a Hubble constant of 50 km s$^{-1}$ Mpc$^{-1}$ and the velocity distance
given by \citet{faber89}, leading to a distance for NGC 4636 of 26.7
Mpc.  A more recent distance determination is given by \citet{tonry01},
where d = 14.7 Mpc for NGC 4636, although \citet{dirs05} argue for a distance of 17.7 Mpc
based on their globular cluster data; here we adopt a distance of 16 Mpc. 
For NGC 5846, we use the distance of 24.9 Mpc obtained by \citet{tonry01}.

\section{Central and Off-Center {\it FUSE\/} Observations of NGC 4636}

\subsection{Observations}

\begin{figure}
\plotone{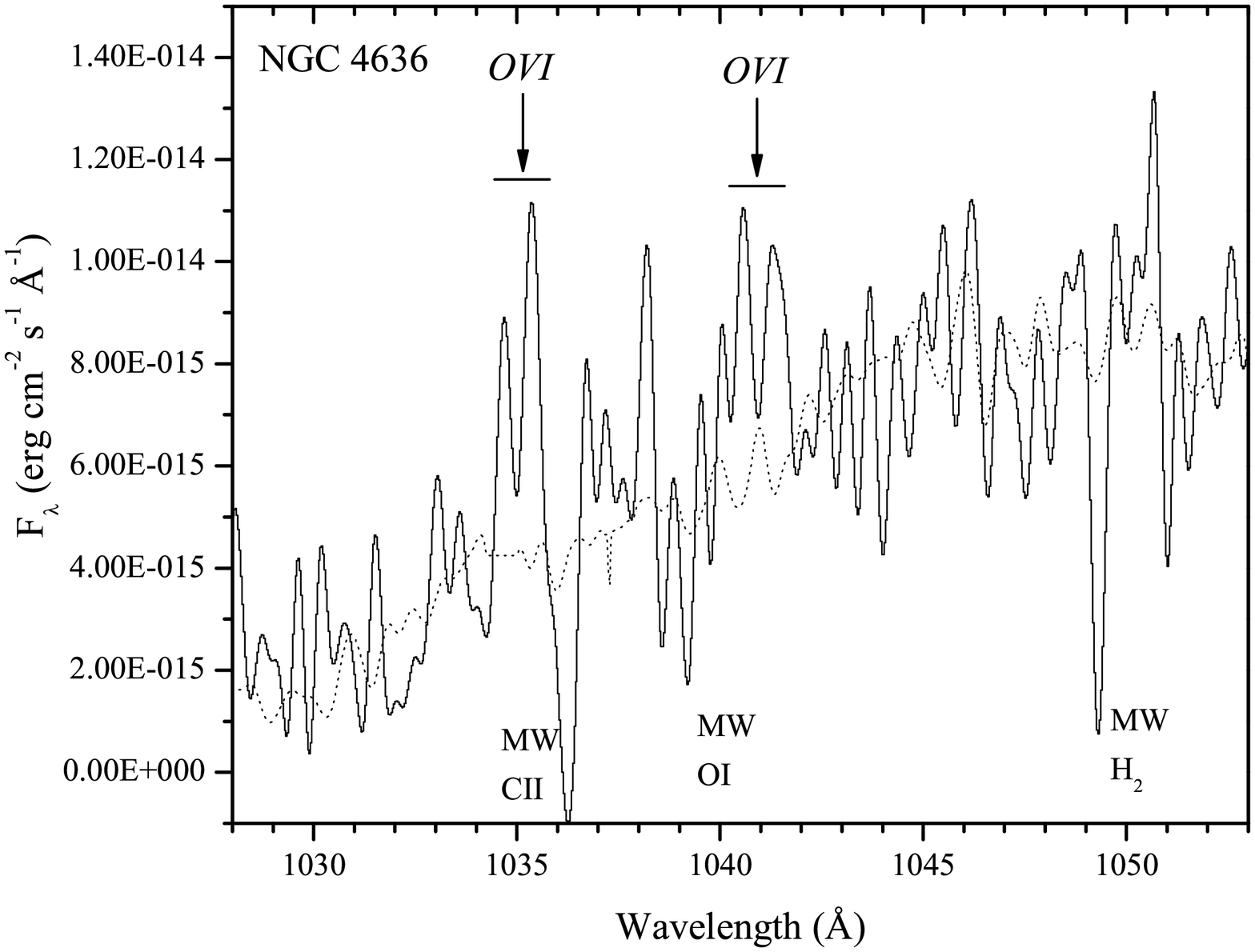}
\caption{The smoothed Lif1a spectrum of NGC 4636 (central position) taken with the LWRS.  The dotted line
is the stellar continuum of the elliptical galaxy NGC 1399, which has a
strong stellar continuum but no \ion{O}{6} emission.  Both strong and weak \ion{O}{6} lines
are detected and are double-peaked with a red and blue component.  Galactic
CII absorption may absorb some of the red side of the strong \ion{O}{6} line.}
\label{fig:n4636spec}
\end{figure}

The first observation, of the central region of NGC 4636 with the
large square aperture (LWRS; 30\arcsec\ square; equivalent in area to a circle of 17\arcsec\ 
radius), was discussed by \citet{breg01}, who reported detections of both
the strong and weak line that were narrow.  Since that time, the pipeline
data processing has improved and led to some changes in the spectrum. 
Originally a wavelength shift needed to be introduced in order for the
Galactic absorption lines to occur at the correct wavelengths, but with
the newer data reduction (versions 2.4 and later of CALFUSE give the
same results), the shift is less than 0.1 \AA\ (30 km s$^{-1}$), which is about as
well as we can measure the lines in this data set.  Also, in \citet{breg01}, both
\ion{O}{6} lines were single-peaked, with the peaks differing in redshift from
the galaxy by about 120 km s$^{-1}$.  With the more recent reduction, which
corrects more precisely for the background, the lines are double-peaked (Figure \ref{fig:n4636spec}),
with the galaxy redshift lying exactly between the two peaks.  Each of the
peaks is relatively narrow, with the strongest feature (the red peak of the
stronger line) being about 0.25 \AA\ wide FWHM (75 km s$^{-1}$); the weaker
features are all about 0.1-0.15 \AA\ wide (all velocities, velocity widths, 
and line strengths are measured from the unsmoothed data).  The separation of the two peaks
for the strong line is about 210 km s$^{-1}$, and this is about the same for the
weaker \ion{O}{6} $\lambda$1038 line.  Also, the line flux for the strong line is about
twice that of the weak line, which is the expected ratio when this doublet
is optically thin.  The flux for the $\lambda$1032 line is 4.0$\times$10$^{{\rm -15}}$ erg cm$^{-2}$ s$^{-1}$ (20\% uncertainty, mainly due to the placement of the continuum) and
for the $\lambda$1038 line it is 2.5 times 10$^{{\rm -15}}$ erg cm$^{-2}$ s$^{-1}$ with similar
uncertainty.  The line centers are at 1035.07 \AA\ and 1040.85 \AA , or
redshifts of 910 km s$^{-1}$ and 933 km s$^{-1}$ (uncertainties of 20 km s$^{-1}$),
which are indistinguishable from the stellar redshift of the galaxy, 938 km s$^{-1}$.

\begin{figure}
\plotone{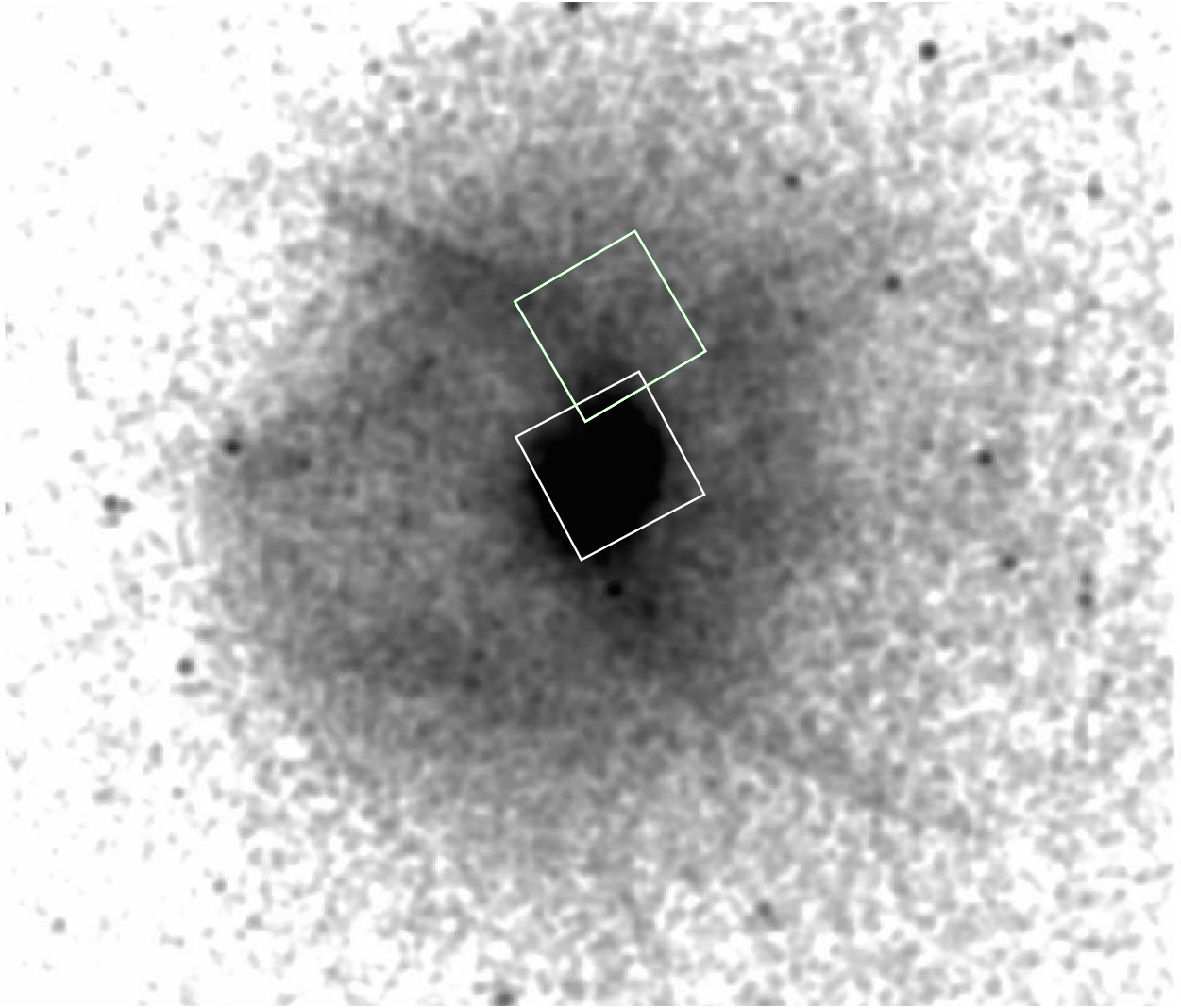}
\caption{The {\em FUSE\/} LWRS apertures (30\arcsec\ square) of the 
central and off-central pointings of NGC 4636, superimposed upon the smoothed Chandra 
0.4--1.5 keV X-ray emission (see a full discussion in \citealt{jones02};
the image is 4\arcmin\ square).}
\label{fig:n4636apertures}
\end{figure}

\begin{figure}
\plotone{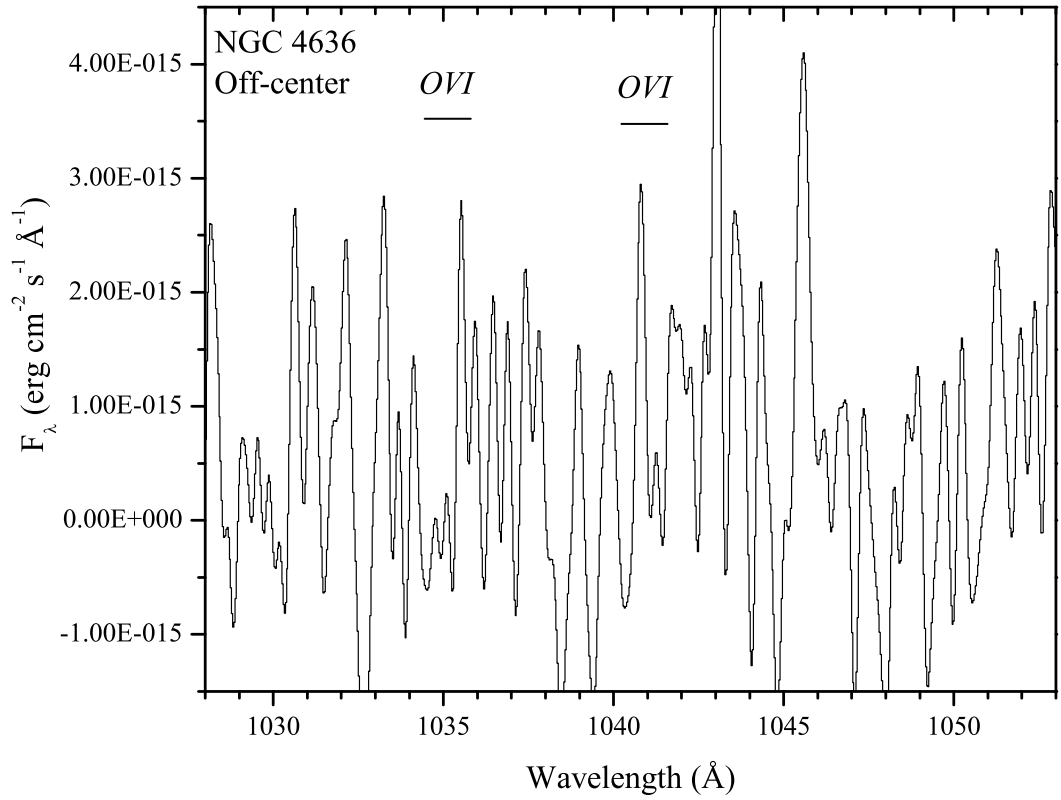}
\caption{The smoothed LWRS Lif1a spectrum of NGC 4636 at the off-center location,
30\arcsec\ north of the galaxy center.  Neither \ion{O}{6} line is detected;
the redshifted line locations are marked.}
\label{fig:n4636spec_offcenter}
\end{figure}

We obtained an off-center pointing with the LWRS for 15.4 ksec in which 13.1
ksec occurred during night, more than twice the observing time of the
central pointing.  The large aperture was placed 30\arcsec\ away from the
center so that the central pointing and the off-center pointing were
adjacent, aside from a slight shift in the position angle (Figure \ref{fig:n4636apertures}).  
This pointing includes emission from r = 1-3 kpc (15-45\arcsec).  
No spectral features were detected (Figure \ref{fig:n4636spec_offcenter}).  
It was expected that the continuum of the galaxy would be too
faint to detect, given optical surface brightness distribution and the
instrumental sensitivity.  Formally, the flux for the stronger \ion{O}{6} $\lambda$1032
line is 0.3$\times$10$^{-16}$ $\pm$ 5$\times$10$^{-16}$ erg cm$^{-2}$ s$^{-1}$, 
so the 3$\sigma$ upper limit is 1.5$\times$10$^{-15}$ erg cm$^{-2}$ s$^{-1}$.

\subsection{Interpretation}

The lack of extended \ion{O}{6} emission constrains a cooling model of 
$q \sim 1$.
In the cooling prescription given above with {\it q\/} near unity, the
cooling rate within an aperture is proportional to the integral of the
surface brightness within the aperture.  The mean X-ray surface
brightness within the off-center aperture is 0.28 times that of the aperture
located at the central region (based on Chandra data with point 
sources subtracted; \citealt{athey05}), so the predicted line strength would have
been F($\lambda$1032) = 1.1$\times$10$^{-15}$ erg cm$^{-2}$ s$^{-1}$.  
This would have been a
2.2$\sigma$ detection had it been present, and as no line is detected, we rule out
the presence of this line at the 98\% confidence level.  The failure to see
this line at the off-center location is consistent with the cooling of the
gas being centrally contained, which supports models with relatively little
mass dropout (e.g., $q \approx\ 0$ models are favored).  In concluding 
that there is a lack of off-center emission, we have assumed that the single 
off-center pointing is representative of other azimuthal locations
at the same radial distance from the center.

The narrow structures of the \ion{O}{6} emission lines also indicate
that the emission line gas is smaller than the 30\arcsec\ square LWRS
aperture.  If the line uniformly filled the aperture, the resolution of the
instrument is reduced to 100 km s$^{-1}$ FWHM (0.34 \AA).  Since we see
\ion{O}{6} line features 0.1-0.25 \AA\ wide, the emitting gas is smaller than the
aperture, approximately by a factor of two.  The observations are
consistent with \ion{O}{6} gas lying within a radius of 10\arcsec.

Under the assumption that all of the \ion{O}{6} emission is contained in
the central pointing, we can calculate a mass cooling rate.  For an
extinction of A$_B$ = 0.12 mag \citep{schleg98} and A(1035 \AA)/A$_V$ =
4.0 \citep{card89}, the luminosity of the strong line is
L($\lambda$1032) = 1.7$\times$10$^{38}$ erg s$^{-1}$.  
For the conversion from L($\lambda$1032) to a mass cooling rate, we use 
the value discussed by \citet{breg01} (see \citealt{edgar86} and \citealt{voit94}),
which leads to $\dot{M}$ = 0.19 $\pm$ 0.04 M$_{\odot}$ yr$^{-1}$.  The cooling rate
predicted from the \citet{sarz89} model gives 0.87 M$_{\odot}$ yr$^{-1}$ 
if we use the X-ray luminosity from the entire galaxy 
and 0.48 M$_{\odot}$ yr$^{-1}$ if
we use the X-ray luminosity from within R$_e$ = 101\arcsec\ ($q = 1$).  The mass
loss rate from stars, based on theoretical stellar evolutionary models is
0.59 M$_{\odot}$ yr$^{-1}$ (see discussion in \citealt{athey02}), 
similar to the X-ray cooling rates but about three times the \ion{O}{6} cooling rate.  These
values are 2.5-4.6 times greater than the cooling rate inferred from the
\ion{O}{6} observations.  To reach consistency between the model and the
observed values, either cooling flows are time dependent or most of the
mass lost by stars does not enter the flow.

The presence of a symmetrical double-peaked
line can be used to rule out spherically symmetric infall.
For an infall model, an optically-thin
spherical shell will have a box-like flat-topped profile that is
twice the width of the infall velocity.  If the emitting volume is
approximated as a series of shells and with a radial velocity gradient,
the emission line will be the sum of flat-topped profiles.
A model such as this will never be double-peaked, so the
\ion{O}{6} emission line data are inconsistent with spherically
symmetric infall.

A rotating disk can produce a double-horned profile, and if
such a structure were present in \ion{O}{6} gas, it might also be apparent in
cooler emission line gas, in dust, or in the stellar surface brightness
distribution.  No central stellar disk is detected in the infrared from
{\it HST NICMOS} observations \citep{ravi01},
and they do not confirm the faint irregular dust extinction
features reported by \citet{vand95}, for which the observations were
obtained with the {\it HST PC} prior to the installation of {\it COSTAR}.  
Optical emission line studies using the H$\alpha$+[NII] lines \citep{buson93,zeil96}
detect 10$^{{\rm 4}}$ K line emission that is distributed over the inner 
10\arcsec\ radius region, similar to the size suggested from the \ion{O}{6}
observations.  The emission line gas is described as being ring-like
because it has a central minimum but then radially decreases in surface
brightness.  Their long-slit spectra show lines with velocity dispersions of
100-200 km s$^{-1}$ (greatest toward the center) and line centers that can
vary by 100 km s$^{-1}$ or more.  The variation in the line center suggests to us 
rotation along several position angles (e.g., P.A. of 85\arcdeg ), so some
of this gas appears to be participating in rotation.  
The velocity structure in the H$\alpha$+[NII] emission lines supports the 
interpretation that the double-horned profile in the \ion{O}{6} gas is
due to rotation.
The separation of the \ion{O}{6} emission peaks implies a projected rotational velocity of
$v_{\rm rot}\sin{i} = $ 105 km s$^{-1}$, which would imply an inclination angle of
$i \approx $ 20\arcdeg.

It seems reasonable to identify this 10$^{{\rm 4}}$ K gas as the
end-product of the \ion{O}{6} emitting material.  The radial distribution of the
10$^{{\rm 4}}$ K gas places it within the location of the \ion{O}{6} gas, and the velocity
properties are similar.  The size of the region with 10$^{{\rm 4}}$ K gas (700 pc) is larger
than that inferred from the $q \approx\ 0$ models (100 pc), a drawback to this
interpretation.  Also, the size of the 10$^{{\rm 4}}$ K gas is similar to the bright
central region seen in X-rays.  The mass of the 10$^{{\rm 4}}$ K gas is highly
uncertain due to the unkown filling factor of the material.

\section{Higher Spatial Resolution Observations of NGC 5846}

\subsection{Observations and Data Reduction}

Since the emission from NGC 4636 seems to be centrally peaked,
we had a program to map the central region of NGC 4636 with the
medium aperture (MDRS; 4\arcsec\ by 20\arcsec), which would also lead to more
accurate line width information.  Unfortunately, loss of a reaction wheel
made scheduling of the observations impossible, so the galaxy NGC
5846 was observed instead.  Although fainter in its \ion{O}{6}
emission, it is a similar galaxy in that it lies in the center of its group and
has optical emission line gas detected.

\begin{figure}
\plotone{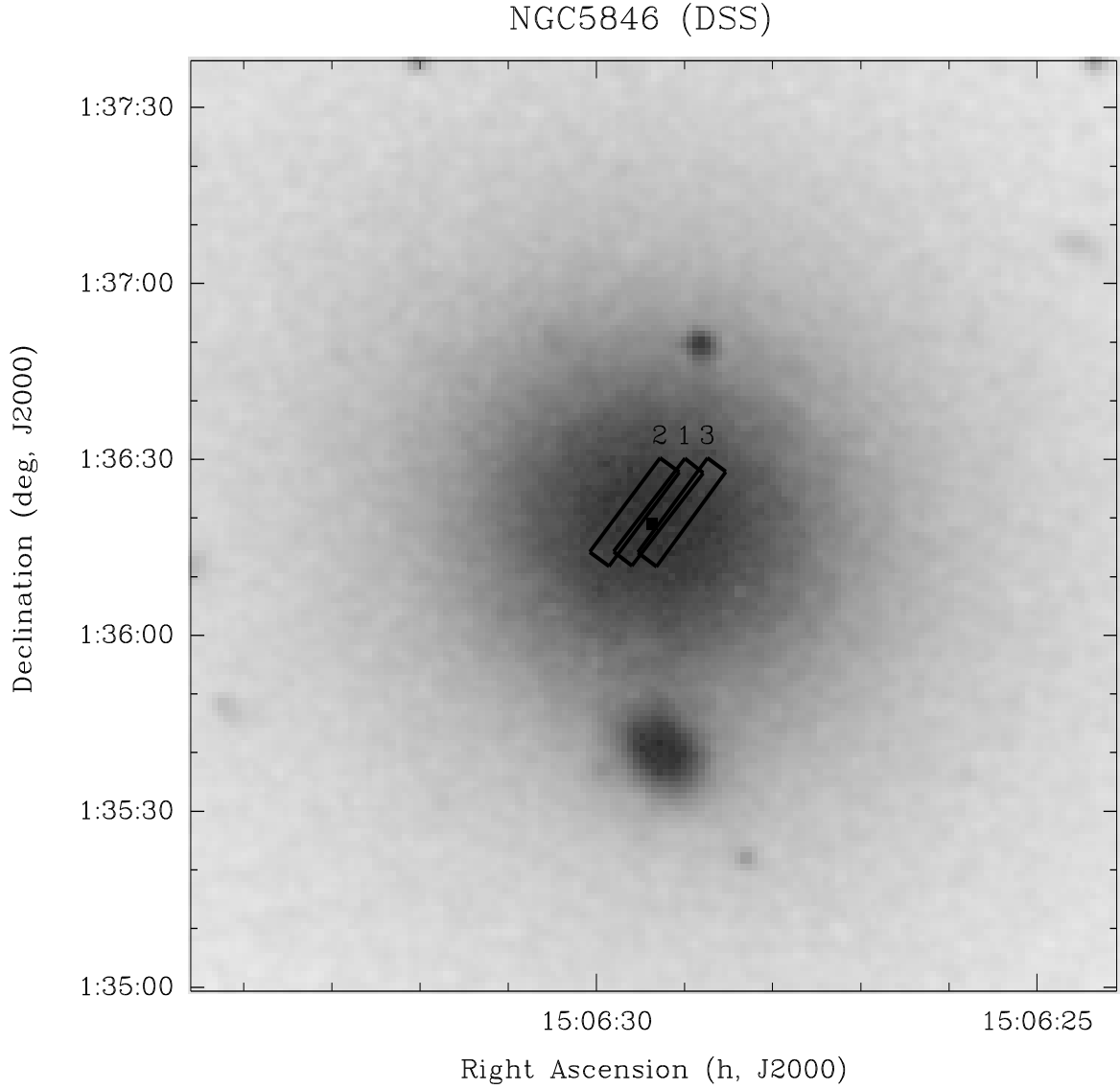}
\caption{The {\em FUSE\/} MDRS apertures superimposed upon the DSS image of
NGC 5846, with the center marked.  The apertures are 4\arcsec\ wide by 20\arcsec\
long.}
\label{fig:n5846apertures}
\end{figure}

The {\em FUSE} observations of NGC\,5846 (IDs C0640101, C0640201, and C0640301) 
were obtained 2004 April 5--6 with the MDRS aperture in timetag mode, with 
exposure times of 21.3, 19.8, and 20.6 ksec, respectively.  The location of the
apertures on NGC 5846 is shown in Figure \ref{fig:n5846apertures}, which have a slight overlap.
The earlier central pointing with the large aperture was for 9.6 ksec 
and is reported upon in \citet{breg05}.

The data reduction and calibration was executed in two different ways in an effort
to optimize the background subtraction and improve the resulting S/N. In the
first method, we combined the exposures of detector segment LiF1A for each
observation using the procedure TTAG\_COMBINE before applying the CALFUSE
pipeline (version 2.4) to each combined observation. We created both background
subtracted and non-background subtracted spectra for day-plus-night and
night-only observations.

In the second method, we applied only the first three steps of the pipeline on
each individual exposure (i.e. initializing the header and computing the Doppler
correction). We then extracted a rectangular region around the \ion{O}{6}
emitting region in the MDRS spectrum and collapsed the extracted region
perpendicular the wavelength dispersion axis to obtain a one-dimensional
spectrum for each exposure. In the case of background subtraction, we extracted
a rectangular region above and below the MDRS aperture and subtracted the
one-dimensional, smoothed average background spectrum from the corresponding
MDRS spectrum. We used the positions of the five \ion{O}{1} airglow lines
(1027.431\,\AA, 1028.157\,\AA, 1039.230\,\AA, 1040.943\,\AA, and 1041.688\,\AA)
surrounding the \ion{O}{6} doublet to derive possible offsets between the
individual exposures. The spectra were then shifted by the corresponding
amounts, before they were combined into one spectrum for each observation. The
\ion{O}{1} airglow lines in the combined spectra were fitted, and their
positions yielded the linear wavelength solution for the small spectral region.
For each observation, a day-plus-night and a night-only spectrum were extracted
using pulse height limits 2--25. The derived wavelength solutions were adjusted
to heliocentric wavelengths. An effective area of 26.5\,cm$^2$ was used to
flux-calibrate the spectra.

\subsection{Measurements}

\begin{figure}
\plotone{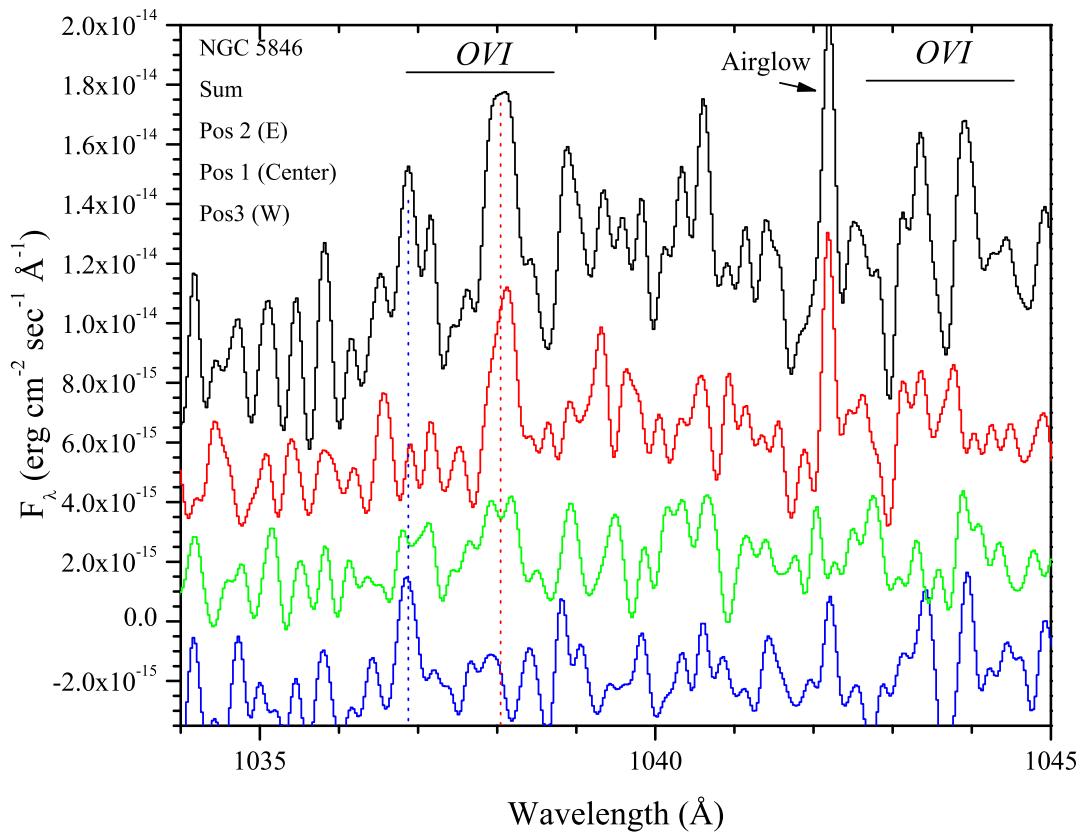}
\caption{The three spectra of NGC 5846, without background subtraction, with 
position 3, 1, 2 (west to east) from the bottom upwards, with the sum of the
spectra at the top; the spectra have been offset for display purposes.  The 
locations of the \ion{O}{6} lines are marked with a bar of length the FWHM of the
velocity dispersion of the galaxy ($\sigma$ = 224 km s$^{-1}$).
The weak line is not detected but the strong line is detected, with both a 
red and blue component associated with the eastern and western sides of the
nucleus (marked with vertical dotted lines).}
\label{fig:n5846spectra}
\end{figure}

The resulting spectra were binned by 8 pixels to improve the signal-to-noise
ratio. A comparison between the day-plus-night and night-only spectra and
between background subtracted and non-background subtracted spectra revealed
that the night-only spectra without background subtraction in general yielded
the better signal-to-noise ratios, so we used these spectra for the analysis. 
The systemic velocity of
NGC\,5846 ($v=1714$\,km\,s$^{-1}$) shifts the \ion{O}{6} $\lambda$1038 emission
line just beyond the \ion{O}{1} airglow lines into a region of scattered light
on the detector. Since this emission line is intrinsically weaker than the
$\lambda$1032 emission line (up to a factor of two), the $\lambda$1038 emission
line became impossible to measure in the spectra in which \ion{O}{6}
$\lambda$1032 was detected (Figure \ref{fig:n5846spectra}).

The \ion{O}{6} $\lambda$1032 emission line was fit in a variety of ways, with
most of the uncertainty due to the placement of the continuum.  
We subtracted a linear fit to the continuum, which includes the background,
to the region 1033-1042 \AA, excluding the line region.  
Other approaches to the line extraction, such as using a constant to fit the
continuum, or using the background-subtracted data provided by the {\it FUSE\/}
pipeline did not lead to changes in the line widths or line centers, but could
cause changes in the line fluxes of 20\%.
Gaussians were fit to the blue and red parts of the line, and the residuals 
are consistent with a normal distribution.

The red component of the redshifted
\ion{O}{6} $\lambda$1032 line has a line center at 1038.07 $\pm$ 0.05 \AA, a line width of 
102 $\pm$ 27 km s$^{-1}$, and a flux of 2.68$\times$10$^{-15}$ erg cm$^{-2}$ s$^{-1}$
$\pm$ 0.69$\times$10$^{-15}$ erg cm$^{-2}$ s$^{-1}$ (a 4$\sigma$ result).  
Nearly all of the flux from this line is due to position 2, and when 
the individual positions are analyzed separately, it is the only location with
a signal above 3$\sigma$ (at 3.2$\sigma$).  The contribution from position 1
is 2.2$\sigma$, and the contribution from position 3 is less than 1$\sigma$.

The weaker blue line only constitutes a 2.5$\sigma$ measurement, it
has a line center at 1036.91 $\pm$ 0.05 \AA, a line width of 
85 $\pm$ 38 km s$^{-1}$, and a flux of 1.56$\times$10$^{-15}$ erg cm$^{-2}$ s$^{-1}$
$\pm$ 0.63$\times$10$^{-15}$ erg cm$^{-2}$ s$^{-1}$.
Nearly all of the contribution to this feature is from position 3, while
the fluxes added from locations 1 and 2 are not statistically significant.
When adding both the blue and red components together, the total 
line flux is 4.24$\times$10$^{-15}$ erg cm$^{-2}$ s$^{-1}$ $\pm$
0.93$\times$10$^{-15}$ erg cm$^{-2}$ s$^{-1}$.
Using the summed spectrum, the red line peak occurs at 
+69$\pm$15\,km\,s$^{-1}$ from the systemic velocity of
NGC 5846, while the blue line peaks at -266$\pm$17\,km\,s$^{-1}$ 
relative to the galaxy.

From the spectra (Figure \ref{fig:n5846spectra}), there is the suggestion 
of a weakening of the red component and a strengthening of the blue component
along the direction east to west.  While this is similar to the structure
of the optical [NII] emission (discussed further below; \citealt{plana98}),
higher signal-to-noise data would be desirable to quantify this trend
independently for the \ion{O}{6} line emission.

\subsection{Interpretation}

One of the goals is to determine the extent of the \ion{O}{6} emission and
the other is to gain more information on the line width, especially in
comparison to the observation taken with the large aperture.
In analyzing the LWRS data \citep{breg05}, we first note that if the \ion{O}{6} 
$\lambda$1032 emission line were at the systemic velocity of NGC 5846, 
it would lie at 1037.8 \AA.  The width and structure of
the line is difficult to determine because Galactic absorption lines modify
the underlying line shape.  The blue side is partly 
absorbed by the strong  Galactic \ion{C}{2} $\lambda$1036.34 absorption line, 
while three Galactic H$_2$ lines absorb the line at 1036.54 \AA\ (on the blue side), 
at 1037.14 \AA\, and at 1038.16 \AA\ (the weakest of the three lines; on
the red side of the \ion{O}{6} line; see \citealt{breg05}).  Nevertheless,
the line appears to have a width of approximately 1 \AA\ FWHM, or 330 km s$^{-1}$
and a line flux of 3.0$\times$10$^{-15}$ erg cm$^{-2}$ s$^{-1}$ before the
Galactic extinction correction. 
To within the uncertainties, the LWRS flux is the same as the line flux measured with the sum
of the MDRS apertures, suggesting that the emission is relatively compact.
 
From the observations with the MDRS aperture, the line strength just at position 2
accounts for most of the total line emission, indicating that most of the emission falls 
within a small region, probably only a few arcseconds in radius and a few
arcseconds away from the nucleus (4\arcsec \ is 480 pc).  
This offset from the nucleus is much larger than the usual pointing 
uncertainty, so we have investigated the uncertainty in the optical positions.  
The optical position of NGC 5846 is based upon the Sloan Digitial 
Sky Survey calibration, which claims an accuracy of 0.5\arcsec\ .  
We conclude that for unknown reasons, the peak in the \ion{O}{6} 
emission is offset from the nucleus.

We find that the \ion{O}{6} line centroids and widths are nearly consistent
with what one would expect if the \ion{O}{6} gas and the 10$^4$ K gas
had the same velocity field.  The velocity field of the 10$^4$ K 
gas that was determined with a scanning Fabry-Perot instrument for the
H$\alpha $ and [NII] $\lambda$6584 lines \citep{plana98}.  They find that the 
[NII] $\lambda$6584 emission is brightest in the nuclear region, decreasing in 
intensity so that it is no longer visible beyond 10-15\arcsec\ from the center.
The line emission is roughly symmetrical, there is a bit more flux to the east,
which is the location of position 2 where we detected the strongest \ion{O}{6} emission.
The velocity field of the 10$^4$ K gas shows a major axis from SE to NW (PA of
130\arcdeg), along which the radial velocity changes from 1820 km s$^{-1}$ in
the SE to 1550 km s$^{-1}$ in the NW (relative to the systemic velocity of 1714 km s$^{-1}$,
this is +105 km s$^{-1}$ in the SE and -164 km s$^{-1}$ in the NW).  
The \ion{O}{6} line width seen in the LWRS observations would be representative of the 
total velocity range.  The \ion{O}{6} line at position 2, which is redshifted
from the systemic velocity of the galaxy by about 69 km s$^{-1}$,
is approximately the velocity one would 
expect from an aperture on the eastern side of the nucleus from the velocity
field of \citet{plana98}.  Also, the range of velocities is only expected to be 
50--100 km s$^{-1}$, similar to the observed line width.
The \ion{O}{6} emission from position 3 (and the blue component of position 1)
is blueshifted from the systemic velocity, as is the [NII] gas, but it is 
more blueshifted than the [NII] gas by about 100 km s$^{-1}$.
This additional blueshift may occur because one of the H$_2$ lines absorbs 
some emission (at 1037.1 \AA) or because there is infall. 
The \ion{O}{6} emission from the central location (position 1) is consistent with both blue
and red emission components, as would be expected since the slit does not
have the same position angle as the rotation axis.

To calculate the luminosity and mass cooling rate, we use an average
of the MDRS and LWRS observations, or F($\lambda$1032) = 
3.6$\times$10$^{-15}$ erg cm$^{-2}$ s$^{-1}$.
After correcting for the Galactic extinction of A$_B$ = 0.237 mag \citep{schleg98},
the flux is 7.1$\times$10$^{-15}$ erg cm$^{-2}$ s$^{-1}$ and the luminosity is
5.2$\times$10$^{38}$ erg s$^{-1}$, or a mass cooling rate of 0.58 M$_{\odot}$ yr$^{-1}$.
It is difficult to estimate the amount of the line removed by the various Galactic
absorption lines, but it is unlikely to add more than 50\% to the line luminosity.
There is very little evidence for dust extinction within NGC 5846 \citep{verd05},
so no corrections of this nature are applied.

The mass flux rate inferred from the X-ray data give values of 1.2 M$_{\odot}$ yr$^{-1}$ 
(0.81 M$_{\odot}$ yr$^{-1}$ within R$_e$), or a factor of 1.4--2.1 above the 
cooling rate inferred from the \ion{O}{6} luminosity.  Given the uncertainty in the
\ion{O}{6} luminosity due to the error and a poorly known Galactic absorption line correction,
as well as model uncertainties, the theoretical and observed cooling rates may be 
consistent with each other.
The stellar mass loss rate for NGC 5846 is about 0.73 M$_{\odot}$ yr$^{-1}$ \citep{athey02}, similar to
the \ion{O}{6} cooling rate.  These data are consistent with a model where the mass
lost from stars becomes thermalized and eventually cools to 10$^4$ K, producing
\ion{O}{6} emission in the process.

\section{Conclusions and Final Comments}

The observations for both NGC 4636 and NGC 5846 indicate a size for the
\ion{O}{6} emission line gas that is probably less than 10\arcsec\ in NGC 4636 (0.8 kpc)
and in the case of NGC 5846, is dominated by a region only 4\arcsec\ in 
size (0.5 kpc).  This compact size rules out cooling flow models where mass drop-out
occurs over a range of radii throughout the galaxy (provided that these off--center
pointings are representative of regions at these radii).  Based on the model of
\citet{sarz89}, the limits on the size of the \ion{O}{6} emission region,
and the ratio of observed to predicted cooling rates, we can rule
out $q \gtrsim\ 1$ for NGC 4636 and $q \gtrsim\ 0.5$ for NGC 5846.  
The line structure and line widths support this picture and suggest
that the \ion{O}{6} emitting gas has either formed a rotating disk or
is in the process of doing so.

The central regions of both galaxies are bright in diffuse X-ray emission, although
as noted for NGC 4636, \ion{O}{7} emission is not detected at luminosities
corresponding to similar values of the cooling rate \citep{xu02}.  Also,
our analysis of the {\it XMM-Newton RGS\/} spectrum of NGC 5846 does not
show the \ion{O}{7} line either (in preparation).  In order to reach consistency with the
\ion{O}{6} emission line data, a turbulent mixing layer would need to occur
that would move the gas quickly from its ambient temperature to below
7$\times$10$^5$ K \citep{slavin93,breg05}.

At the rate that the gas is cooling, it cannot accumulate for more than
10$^7$--10$^8$ yr or a significant amount of neutral gas would 
be detectable as 21 cm emission, in conflict with upper limits
\citep{roberts91}.  Either this gas feeds a black hole, causing intermittent
behavior of an active galactic nucleus, or the gas is consumed in star formation.
Both galaxies have similar 1.4 GHz radio luminosities (21 mJy for NGC 5846 and
78 mJy for NGC 4636; \citealt{condon98}), which may suggest the presence of some
type of AGN activity, and there may be evidence for an
interaction between a radio jet and the hot ambient medium \citep{jones02}.
Star formation, if present, has eluded previous efforts to detect it, but
observations with the {\it Spitzer\/} observatory may prove more definitive.

\acknowledgements
We would like to thank the {\it FUSE\/} team for their assistance in the 
collection and reduction of these data, and George Sonneborn and B-G 
Andersson for rescheduling the target in program C064.  
This research has made use of the NASA/IPAC Extragalactic Database (operated by JPL, Caltech),
the Multimission Archive at Space Telescope, and the NASA Astrophysics 
Data System, operated under contract with NASA.
We gratefully acknowledge support by NASA through grants NAG5-9021, 
NAG5-11483, G01-2089, GO1-2087, GO2-3114, and NAG5-10765.

\clearpage

\begin{thebibliography}{}

\bibitem[Athey et al.(2002)]{athey02} Athey, A., Bregman, J.N., Bregman, J.D., Temi, P., \& Sauvage, M. 2002, \apj, 571, 272
\bibitem[Athey, Bregman, and Irwin(2005)]{athey05}  Athey, A.E., Bregman, J.N., \& Irwin, J.A. 2005, \apjs, submitted
\bibitem[Balbus(1988)]{balb88} Balbus, S.~A.\ 1988, \apj, 328, 395
\bibitem[Balbus(1991)]{balb91} Balbus, S.~A.\ 1991, \apj, 372, 25 
\bibitem[Balbus \& Soker(1989)]{balb89} Balbus, S.~A., \& Soker, N.\ 1989, \apj, 341, 611
\bibitem[Bregman et al.(2005)]{breg05} Bregman, J.N., Miller, E.D., Athey, A.E.,
\& Irwin, J.A. 2005, \apj, submitted
\bibitem[Bregman, Miller, \& Irwin(2001)]{breg01}  Bregman, J.N., Miller, E.D., \& Irwin, J.A. 2001, \apjl, 553, L125
\bibitem[Buson et al.(1993)]{buson93} Buson, L.~M., et al.\ 
1993, \aap, 280, 409
\bibitem[Cardelli et al.(1989)]{card89} Cardelli, J.~A., Clayton, G.~C., \& Mathis, J.~S.\ 1989, \apj, 345, 245
\bibitem[Condon et al.(1998)]{condon98} Condon, J.~J., Cotton, W.~D., Greisen, 
E.~W., Yin, Q.~F., Perley, R.~A., Taylor, G.~B., \& Broderick, J.~J.\ 1998, \aj, 115, 1693
\bibitem[Dirsch, Schuberth, \& Richtler(2005)]{dirs05} Dirsch, B., Schuberth, 
Y., \& Richtler, T.\ 2005, \aap, 433, 43
\bibitem[Edgar \& Chevalier(1986)]{edgar86}  Edgar, R.J., \& Chevalier, R.A. 1986, 310, L27
\bibitem[Faber et al.(1989)]{faber89}  Faber, S.~M., Wegner, G., 
Burstein, D., Davies, R.~L., Dressler, A., Lynden-Bell, D., \& Terlevich, 
R.~J.\ 1989, \apjs, 69, 763 
%%\bibitem[Frei et al.(1996)]{frei96} Frei, Z., Guhathakurta, 
P., Gunn, J.~E., \& Tyson, J.~A.\ 1996, \aj, 111, 174
\bibitem[Jones et al.(2002)]{jones02} Jones, C., Forman, W., 
Vikhlinin, A., Markevitch, M., David, L., Warmflash, A., Murray, S., \& 
Nulsen, P.~E.~J.\ 2002, \apjl, 567, L115
\bibitem[Mathews \& Bregman(1978)]{math78} Mathews, W.~G., \& 
Bregman, J.~N.\ 1978, \apj, 224, 308
\bibitem[Mathews \& Brighenti(2003)]{math03} Mathews, W.G., \& Brighenti, F. 2003, \araa, 41, 191
\bibitem[Plana et al.(1998)]{plana98} Plana, H., Boulesteix, J., Amram, P., 
Carignan, C., \& Mendes de Oliveira, C.\ 1998, \aaps, 128, 75
\bibitem[Ravindranath et al.(2001)]{ravi01} Ravindranath, S., 
Ho, L.~C., Peng, C.~Y., Filippenko, A.~V., \& Sargent, W.~L.~W.\ 2001, \aj, 
122, 653
\bibitem[Roberts et al.(1991)]{roberts91} Roberts, M.S., Hogg, D.E., Bregman, J.N., Forman, W.R., \& Jones, C. 1991, \apjs, 75, 751
\bibitem[Sarazin \& Ashe(1989)]{sarz89} Sarazin, C.L. \& Ashe, G.A. 1989, \apj, 345, 22
%%\bibitem[Sarazin \& White (1987)], {sarz87} Sarazin, C.L. \& White, R.E 1987, \apj 320, 32
\bibitem[Schlegel, Finkbeiner, \& Davis(1998)]{schleg98}  Schlegel, D.J., Finkbeiner, D.P., \& Davis, M. 1998, \apj, 500, 525
\bibitem[Slavin, Shull, \& Begelman(1993)]{slavin93} Slavin, J.D., Shull, J.M., \& Begelman, M.C. 1993, \apj, 407, 83
\bibitem[Tonry et al.(2001)]{tonry01} Tonry, J.~L., Dressler, 
A., Blakeslee, J.~P., Ajhar, E.~A., Fletcher, A.~B., Luppino, G.~A., 
Metzger, M.~R., \& Moore, C.~B.\ 2001, \apj, 546, 681
\bibitem[van Dokkum \& Franx(1995)]{vand95} van Dokkum, P.~G., 
\& Franx, M.\ 1995, \aj, 110, 2027
\bibitem[Verdoes Kleijn \& de Zeeuw(2005)]{verd05} Verdoes 
Kleijn, G.~A., \& de Zeeuw, P.~T.\ 2005, \aap, 435, 43 
\bibitem[Voit, Donahue, \& Slavin(1994)]{voit94} Voit, G.M., Donahue, M., \& Slavin, J.D. 1994, \apjs, 95, 87
\bibitem[Xu et al.(2002)]{xu02} Xu, H., et al.\ 2002, \apj, 
579, 600
\bibitem[Zeilinger et al.(1996)]{zeil96} Zeilinger, W.~W., et 
al.\ 1996, \aaps, 120, 257

\end{thebibliography}
\end{document}